\documentclass[%
superscriptaddress,
preprint,
amsmath,amssymb,
aps,
]{revtex4-1}

\usepackage[utf8]{inputenc}
\usepackage{graphicx}
\usepackage{dcolumn}
\usepackage{bm}
\usepackage{xcolor}

\usepackage{lineno}
\newcommand*\patchAmsMathEnvironmentForLineno[1]{%
	\expandafter\let\csname old#1\expandafter\endcsname\csname #1\endcsname
	\expandafter\let\csname oldend#1\expandafter\endcsname\csname end#1\endcsname
	\renewenvironment{#1}%
	{\linenomath\csname old#1\endcsname}%
	{\csname oldend#1\endcsname\endlinenomath}}%
\newcommand*\patchBothAmsMathEnvironmentsForLineno[1]{%
	\patchAmsMathEnvironmentForLineno{#1}%
	\patchAmsMathEnvironmentForLineno{#1*}}%
\AtBeginDocument{%
	\patchBothAmsMathEnvironmentsForLineno{equation}%
	\patchBothAmsMathEnvironmentsForLineno{align}%
	\patchBothAmsMathEnvironmentsForLineno{flalign}%
	\patchBothAmsMathEnvironmentsForLineno{alignat}%
	\patchBothAmsMathEnvironmentsForLineno{gather}%
	\patchBothAmsMathEnvironmentsForLineno{multline}%
}

\usepackage{color}

\begin{document}

\title{Multi frequency matching for voltage waveform tailoring}

\author{Frederik Schmidt}
\affiliation{Ruhr University Bochum, 44780 Bochum, Germany}
\author{Julian Schulze}
\affiliation{Ruhr University Bochum, 44780 Bochum, Germany}
\affiliation{Department of Physics, West Virginia University, Morgantown, West Virginia 26506-6315, USA}
\author{Erik Johnson}
\affiliation{LPICM-CNRS, Ecole Polytechnique, 91128 Palaiseau, France}
\author{Jean-Paul Booth}
\affiliation{LPP, (CNRS, Université Pierre et Marie Curie, Université Paris-Sud, Université Paris-Saclay, Sorbonne Universités), Ecole Polytechnique, Route de Saclay, 91128 Palaiseau,
France}
\author{Douglas Keil}
\affiliation{Lam Research Corp., Tualatin, Oregon 97062, USA}
\author{David M. French}
\affiliation{Lam Research Corp., Tualatin, Oregon 97062, USA}
\author{Jan Trieschmann}
\affiliation{Brandenburg University of Technology, Electrodynamics and Physical Electronics Group, 03046 Cottbus, Germany}
\author{Thomas Mussenbrock}
\affiliation{Brandenburg University of Technology, Electrodynamics and Physical Electronics Group, 03046 Cottbus, Germany}

\date{\today}

\begin{abstract}
Customized voltage waveforms composed of a number of frequencies and used as the excitation of radio-frequency plasmas can control various plasma parameters such as energy distribution functions, homogeneity of the ionflux or ionization dynamics. So far this technology, while being extensively studied in academia, has yet to be established in applications. One reason for this is the lack of a suitable multi-frequency matching network that allows for maximum power absorption for each excitation frequency that is generated and transmitted via a single broadband amplifier. In this work, a method is introduced for designing such a network based on network theory and synthesis. Using this method, a circuit simulation is established that connects an exemplary matching network to an equivalent circuit plasma model of a capacitive radio-frequency discharge. It is found that for a range of gas pressures and number of excitation frequencies the matching conditions can be satisfied, which proves the functionality and feasibility of the proposed concept. Based on the proposed multi-frequency impedance matching, tailored voltage waveforms can be used at an industrial level.

\end{abstract}

\maketitle


\section{\label{sec:introduction} Introduction}

Control of energy distribution functions of different particle species (electrons, ions, neutrals) is crucially important to optimize various applications of low temperature technological radio frequency (RF) plasmas ranging from solar cell manufacturing to biomedicine and integrated circuits \cite{lieberman_principles_2005,chabert_physics_2011,samukawa_2012_2012}. Based on the pioneering work of Wendt et al. \cite{patterson_arbitrary_2007,wang_control_2000} and the discovery of the Electrical Asymmetry Effect (EAE) \cite{heil_numerical_2008,heil_possibility_2008,donko_pic_2009,schulze_electrical_2009,schungel_electrical_2012,schungel_power_2011,schungel_effect_2013}, a new concept to realize such advanced process control has been developed. It is based on driving RF plasmas with customized voltage waveforms generated as a finite Fourier Series of multiple consecutive harmonics of a fundamental frequency with individually adjustable harmonic amplitudes and phases \cite{lafleur_tailored-waveform_2016,schulze_electrical_2011,lafleur_separate_2012,diomede_radio-frequency_2014,berger_experimental_2015,derzsi_electron_2013,delattre_radio-frequency_2013,bruneau_ion_2014,bruneau_control_2014,bruneau_strong_2015,schungel_customized_2015}. In capacitive RF plasmas, this so called Voltage Waveform Tailoring (VWT) has proven to allow customization of the electron heating dynamics and the sheath voltage as a function of time within the fundamental RF period. It has been found that the symmetry of the plasma can be controlled by tuning the relative phases of the  harmonics, either through the electrical generation of a DC self bias when there is an amplitude asymmetry in the driving voltage waveform  or by inducing different ionization dynamics adjacent to each electrode with a slope asymmetry in the driving voltage waveform. These effects allow the electron and ion energy distribution functions to be customized in the plasma volume and at boundary surfaces. Note that the control of the ion energy distribution depends on the driving frequencies. While at low driving frequencies the ions can typically react to the time dependent electric field in the sheaths, they cannot react at higher driving frequencies. Nevertheless, previous works demonstrated that VWT allows to control these distribution functions in both frequency domains \cite{patterson_arbitrary_2007,wang_control_2000,schulze_electrical_2011,schungel_customized_2015}. At high driving frequencies, the mean ion energy can be adjusted by phase control via the EAE. Even the shape of the IEDF can be controlled in this case by customizing the sheath voltage waveform in a way that allows low energy ions generated by charge exchange collisions inside the sheath to be accelerated to distinct energies \cite{schungel_prevention_2015}. The opportunity to control the electron energy distribution in turn allows control of the generation of reactive neutrals, and therefore their densities. At high driving frequencies electromagnetic effect such as the standing wave effect can cause significant and unwanted lateral non-uniformities of the ion and radical flux across large wafers. Schuengel et al. \cite{schungel_prevention_2015} demonstrated that such lateral non-uniformities of the ion flux can be prevented by VWT. Iwashita et al. \cite{iwashita_transport_2013,iwashita_sheath--sheath_2012} showed that this technology can be used to control the transport of dust particles in such discharges. Recently, VWT has also been applied to inductive RF discharges with phase-locked RF substrate bias and has been demonstrated to provide unique advantages for optimized process control \cite{ahr_influence_2015,berger_enhanced_2017}.

Based on these fundamental insights, VWT has been demonstrated to provide key advantages for various plasma processes: Johnson et al. \cite{johnson_nanocrystalline_2010,johnson_microcrystalline_2012,johnson_hydrogenated_2012} and Hrunski et al. \cite{hrunski_influence_2013,hrunski_deposition_2013} showed that the deposition rate and characteristics of Si:H thin films deposited by Plasma Enhanced Chemical Vapor Deposition (PECVD) in capacitive RF plasmas could be optimized, controlled, and their uniformity could be improved by VWT. Wang et al. \cite{wang_electrode-selective_2016} showed that VWT allows switching from etching to deposition in the same reactor by changing the driving voltage waveform from sawtooth up to sawtooth down. Zhang et al. \cite{zhang_control_2015} showed that VWT and phase control are beneficial for plasma etching as well.  

All these significant enhancements of process control can be achieved without mechanical modification of the plasma reactor. Only the external circuit (the RF generators and impedance matching network)  needs to be adapted. In this sense, VWT is a modular technique that can simply be added to any existing RF plasma chamber as an upgrade.

To date most (fundamental and applied) studies of VWT have been performed on an academic level, i.e., in small to medium scale reactors and using relatively simple plasma chemistries at low powers; however they have demonstrated enormous potential to improve industrial plasma processing applications typically performed at much higher powers, in complex gas mixtures, and in large reactors. The upscaling of this technology to industrial standards is hindered by the lack of cost-effective concepts for impedance matching of such complex multi-frequency discharges. Such impedance matching is required, since an RF plasma represents a complex load to an RF generator, which typically has an output impedance of 50 Ohm without any imaginary component; at low pressure, a capacitive RF plasma behaves predominantly as a capacitive load with a large imaginary component of its impedance. Therefore a matching network is required to match the impedance of the plasma load to the impedance of the generator to maximize the power dissipated in the plasma \cite{lieberman_principles_2005,chabert_physics_2011}. Poor impedance matching results in high reflected powers, which can damage or even destroy the RF generator(s). While impedance matching circuits exist for single or dual frequency power sources, concepts for such networks for multi-frequency tailored voltage waveforms are strongly limited, since the load impedance is frequency dependent and matching must be achieved for all driving frequencies simultaneously. As a consequence, classical matching networks cannot be used. 

The only design concept that has been demonstrated for an impedance matching for tailored voltage waveforms has been proposed by Franek et al. \cite{franek_power_2015}. It is based on separate generation of each harmonic of a given multi-frequency waveform. Each harmonic is then impedance matched by a separate matching network, and subsequently all harmonics are combined to drive the electrode in a capacitively coupled RF discharge. This concept is based on using multiple conventional single-frequency impedance matchings in multiple matching branches each protected by band-pass filters to prevent parasitic coupling between the individual matching branches. While this concept is fully functional \cite{franek_power_2015,berger_experimental_2015}, it is not an efficient solution to the problem of generating  impedance-matched tailored voltage waveforms, because it does not allow the generation of these waveforms by the combination of an arbitrary function generator and a single broadband amplifier. This, however, would be the ideal and most cost-effective approach at an industrial level, particularly if a high number of harmonics is required. Therefore, a novel concept for multi-frequency impedance matching of VWT is desired, which does not require separate matching branches for each harmonic and which can be used in combination with a single broadband amplifier with a single input line and a single output line which can be directly connected to the plasma source.

Here, we present a solution to this problem, i.e., we propose a method for matching a frequency-dependent load to a multi-frequency RF generator via a single line and without the need for multiple generators and filternetworks. This concept is developed based on network theory and a global plasma model coupled self-consistently to a customized external circuit. In this way, a network consisting of inductances and capacitances (including losses and stray effects) is proposed, which is able to transform the load impedance to the internal resistance of the generator for each applied frequency and, therefore, allows for maximum power transfer from the generator to the load in the presence of multiple driving frequencies. To prove the feasibility of this new multi-frequency impedance matching concept, exemplary simulations of a capacitively coupled plasma are carried out for different neutral gas pressures and driving voltage waveforms. In each scenario, excellent impedance matching is achieved for all driving frequencies.

The paper is structured in the following way: In section 2, the new concept for impedance matching for VWT is introduced. In section 3, the simulation model and the algorithm used to test the matching performance with a capacitive plasma reactor are outlined. Illustrative scenarios of such discharges driven by 3 and 5 consecutive harmonics of 13.56 MHz at low and high neutral gas pressures in argon are studied in section 4, where the performance of the impedance matching is discussed in terms of forwarded and reflected powers. Finally, conclusions are drawn in section 5.


\section{\label{sec:method} Matching Method}

\begin{figure}[t!]
	\centering
	\resizebox{12cm}{!}{		
		\includegraphics[width=8cm]{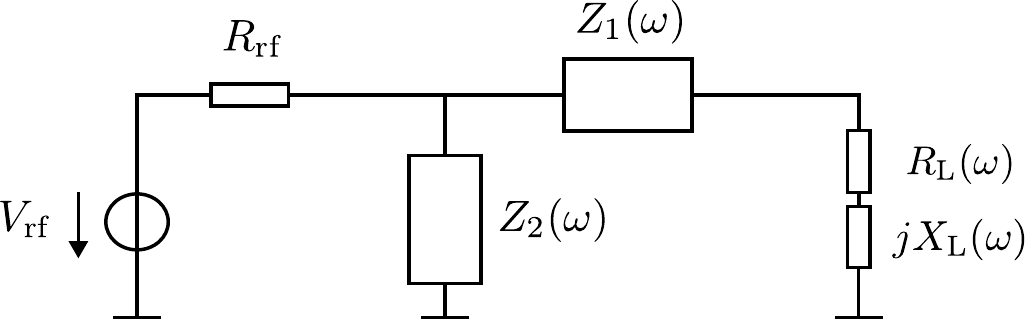}
	}
	\caption{Voltage source with internal resistance attached to a frequency dependent complex load via a matching network consisting of two impedances $Z_1$ and $Z_2$.}
	\label{fig:introduction}
\end{figure}

\begin{figure}[t!]
	\centering
	\resizebox{12cm}{!}{		
		\includegraphics[width=8cm]{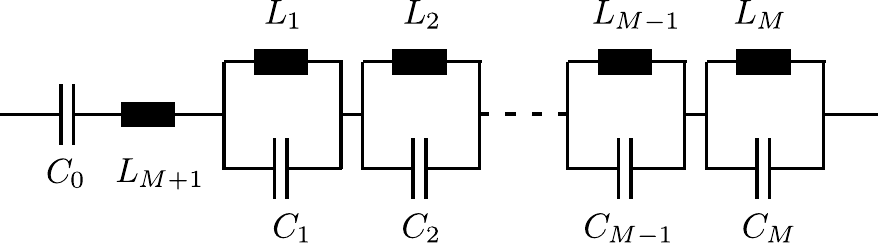}
	}
	\caption{General network consisting of inductances and capacitances following the transformation function for LC-networks.}
	\label{fig:matching_system}
\end{figure}

In order to transfer maximal power from a generator to a load, the load impedance must be transformed to match the internal resistance of the generator. This is typically achieved using impedance matching networks. Following the argument of Roy~\cite{dutta_roy_triple_2014}, first, we generally consider the setup shown in figure~\ref{fig:introduction}. The generator consists of an internal resistance $R_\mathrm{rf}$ and a voltage source of the form $V_\mathrm{rf} = \sum_{i=1}^{N} V_i \mathrm{cos}(\omega_i t + \varphi_i)$, with the amplitude $V_i$, angular frequency $\omega_i$ and phase shift $\varphi_i$ of the $i$th frequency component and $N$ being the total number of frequency components. The absorbed power in the load is maximal, when its resistive part is $R_\mathrm{L}=R_\mathrm{rf}$ and its imaginary part is $X_\mathrm{L}=0$. Since this is generally not the case, an L-network with the impedances $Z_1$ and $Z_2$ is added consisting of inductances and capacitances, whereby we can write $Z_1 = j X_1$ and $Z_2 = j X_2$. To satisfy the matching condition for the real part Re$\{R_\mathrm{rf} || j X_2\} = R_\mathrm{L}$ is required for every excitation frequency. This leads to
\begin{align}
X_2 = \pm \frac{R_\mathrm{rf} R_\mathrm{L}}{\sqrt{R_\mathrm{L}(R_\mathrm{rf}-R_\mathrm{L})}},
\end{align}
with the condition $R_\mathrm{rf}>R_\mathrm{L}$. Since $R_\mathrm{L}=R_\mathrm{L}(\omega)$ is a function of the excitation frequency, we can specify $N$ conditions
\begin{align}
X_{\mathrm{2},i}= \pm \frac{R_\mathrm{rf} R_{\mathrm{L},i}}{\sqrt{R_{\mathrm{L},i}(R_\mathrm{rf}-R_{\mathrm{L},i})}}.
\label{eq:real_part}
\end{align} 
Therein, double indices denote the corresponding impedance and the respective frequency component (e.g., $R_{\mathrm{L},i} \equiv R_\mathrm{L}(\omega_i)$ with $i \in \left[ 1,N \right]$). For $Z_1$, similar considerations for the imaginary part lead to the condition $X_1 = - \mathrm{Im}\{R_\mathrm{rf} || j X_2   \} -X_\mathrm{L}$. This results in $N$ additional equations of the form
\begin{align}
 X_{\mathrm{1},i}= -\frac{R_\mathrm{rf}^2 X_{\mathrm{2},i}}{R_\mathrm{rf}^2 + X_{\mathrm{2},i}^2} - X_{\mathrm{L},i}.
\label{eq:im_part}
\end{align} 
For both $X_1$ and $X_2$, networks consisting of inductances and capacitances must be designed such as to have the values defined in equation systems~\eqref{eq:real_part} and \eqref{eq:im_part}  at the different excitation frequencies. Several methods exist to synthesize LC-networks, e.g., a ladder structure called \textit{Cauer} ladder network~\cite{cauer_verwirklichung_1926} or a \textit{Foster}-type network that can be described by a series of fractions~\cite{foster_reactance_1924}. We make use of the latter case, which allows a transformation function $S$ to be set up in frequency space in the form of 
\begin{align}
S = \frac{k_0}{j \omega} + j \omega k_{M+1}  + \sum_{m=1}^M \frac{j \omega k_m}{q_m - \omega^2},
 \label{eq:transformation}
\end{align}
where the $k$'s and the $q$'s are $2(M+1)$ independent network variables \cite{kuo_network_1966}. $S$ can either be an impedance or an admittance. In this work, we only use $S$ as an impedance, which leads to a network of elements in series described by the different terms in equation~\eqref{eq:transformation}: The first term can be interpreted as a capacitance with $k_0 = 1/C_0$, the second term as an inductance with $k_{M+1} = L_{M+1}$, and the last term as $M$ parallel resonance circuits with $k_m = 1/C_m$ and $q_m = 1/(C_m L_m)$. The resulting network is depicted in figure~\ref{fig:matching_system}. Using this method to build the networks for $X_1$ and $X_2$ respectively, equations \eqref{eq:real_part} and \eqref{eq:im_part} lead to two systems of equations consisting of $N$ equations each, that can be solved for the unknown $k$'s and $q$'s, which in return provide the values of the inductances and capacitances.

To determine how many unknowns are necessary in order to be able to match an arbitrary load, we consider that a general reactance always has a positive slope with increasing $\omega$ \cite{kuo_network_1966}. This fact can  be seen directly by investigating the different terms of equation~\eqref{eq:transformation}, which all have a positive slope with increasing $\omega$. This consequently applies as well to the sum of all individual terms. Each of the $N$ right hand sides of equations \eqref{eq:real_part} and \eqref{eq:im_part} define the specific values the respective networks must have at each excitation frequency. If now these values decrease with increasing $i$, a reactance with a \textit{negative} frequency slope results. For $X_1$ and $X_2$ to be able to still satisfy the equations, a certain number of intermediate poles and zeros in $X_1$ and $X_2$ are required. For $N$ such right hand side values (being either all positive or all negative), the networks must have either $N$ poles or $N$ zeros in $0< \omega < \infty $. The first case translates to $N$ LC-parallel units, or in other words, $k_0 = 0$, $k_{M+1} = 0$ and $M = N$. In the second case, $k_0$ and $k_{M+1}$ are both nonzero and $M = N-1$ LC-parallel units are required (cf: figure~\ref{fig:setup} and \ref{fig:network2}).

This worst-case scenario, with a right hand side of negative slope, can be avoided by taking advantage of the freedom of sign in equations~\eqref{eq:real_part}. Therewith, the number of required elements reduces to the number of frequencies to be matched. In other words, $N$ independent variables in $S$ are necessary to build the network, e.g., for an even $N$ set $k_0 = 0$, $k_{M+1} =0$ and $M = N/2$. This is, however, not feasible in the scenario investigated in this work: If the considered load is nonlinear, as a plasma in fact is, the matching is typically done during process by using tunable capacitances, since tunable inductances for high powers are difficult to construct. In this case, all degrees of freedom must be achieved by varying the capacitances, and the values of the inductances must be chosen beforehand and fixed during the process. Therefore, $N$ capacitances must be used in the network, satisfying equations~\eqref{eq:real_part}. Consequently, in the proposed network $k_{M+1} =0$, $k_0=0$ and $M=N$, resulting in $M$ LC-parallel units in series. Another option would be to choose $k_{M+1}$ and $k_0$ to be nonzero and $M=N-1$. Notice that both options are the same as described in the previous paragraph, and, therefore, the same network setup is used for both $X_1$ and $X_2$. 

To deduce the required values of the elements in the networks, the systems of equations for $X_1$ and $X_2$ must be solved for the network variables $k_m$ and $q_m$. Certain constraints may be added, since there are two times more independent variables than equations. As pointed out before, it is meaningful to choose the values of the inductances beforehand and, thereby, reduce the number of unknowns to the number of equations.

In reality, inductances and capacitances are not ideal elements, but certain stray effects and resistive losses will be present. Therefore, depending on how influential these effects are, the calculations presented in this chapter must be adjusted. To account for resistive losses, $Z_1$ and $Z_2$ must include a frequency-dependent real part. Since stray effects can be represented by inductances and capacitances, and all networks consisting of inductances and capacitances can be transformed in the way described by equation~\eqref{eq:transformation}, the argument in this chapter is still valid. However, the specific values of the elements may need to be adjusted in order to achieve a given impedance at the generator. Both stray effects and losses can be included in theoretical investigations and simulations, but the exact values are often not known in advance. Therefore, we suggest for a specific design in practice to build the matching network based on the presented considerations and measure the impedance function of $Z_1$ and $Z_2$ using a network analyzer beforehand or using  a V-I-probe during process. This way the deviations from the preliminary theoretical  model can be calculated and the elements of the network adjusted accordingly. This approach may fail however, if the strayeffects and losses in the network are too large, resulting in impedances that may not be able to satisfy equations~\eqref{eq:real_part} and~\eqref{eq:im_part}. This problem is best avoided by choosing capacitances and inductances that are as ideal as possible, without the stray effects or losses dominating their properties.

\section{\label{sec:model} Simulation Model and Algorithm}

In order to investigate the performance of the proposed matching method for a simulated plasma, a suitable plasma model must be chosen, which can easily be simulated simultaneously and self-consistently with an attached network consisting of lumped elements. An obvious choice here is an equivalent circuit model, so the problem reduces to a circuit simulation. A well-studied nonlinear equivalent circuit model of a capacitively coupled plasma (CCP) based on the considerations in~\cite{mussenbrock_enhancement_2008,mussenbrock_nonlinear_2006,lieberman_effects_2008,ziegler_temporal_2009,mussenbrock_nonlinear_2007} is used in this work and briefly discussed in the subsequent paragraphs. A more detailed description of the simulation model and algorithm used in this work can be found elsewhere~\cite{schmidt_consistent_2018}.

Following a generalized Ohm's law of the form $\partial_t \vec{j}=e^2 n \vec{E}/m_\mathrm{e} - \nu_\mathrm{eff} \vec{j}$, the bulk can be modeled by a resistance $R_\mathrm{pl}$ and an inductance $L_\mathrm{pl}$, which mimic inelastic collisions and electron inertia, respectively. $n$ is the plasma density, $m_\mathrm{e}$ the electron mass, $\vec{j}$ the conduction current density, and $\vec{E}$ the electric field. $\nu_\mathrm{eff}$ includes both ohmic and stochastic heating, which is obtained from the expression $\nu_\mathrm{eff} = \nu_\mathrm{m} + \bar{v}_\mathrm{e}/l_\mathrm{B}$, with $\nu_\mathrm{m}$ the momentum transfer collision frequency, the mean thermal speed  $\bar{v}_\mathrm{e} = \left( 8 k_\mathrm{B} T_\mathrm{e}/\pi m_\mathrm{e} \right)^\frac{1}{2}$, the electron temperature $T_\mathrm{e}$, the effective bulk length $l_\mathrm{B}$, and the Boltzmann constant $k_\mathrm{B}$. The sheath on the other hand is modeled by a capacitive diode, with a nonlinear sheath capacitance, a constant ioncurrent and a varying electron current. This leads to the system of differential equations
\begin{eqnarray}
 \frac{d V_\mathrm{S,1}}{d t} &=& -C_\mathrm{S,1}^{-1}(I_\mathrm{pl} + j_\mathrm{i0} A_\mathrm{E} - j_\mathrm{e0} A_\mathrm{E} \exp{\left(- \frac{ e V_\mathrm{S,1}}{k_\mathrm{B} T_\mathrm{e}}\right)}, \\
 \frac{d V_\mathrm{S,2}}{d t} &=& -C_\mathrm{S,2}^{-1}(-I_\mathrm{pl} + j_\mathrm{i0} A_\mathrm{G} - j_\mathrm{e0} A_\mathrm{G} \exp{\left(- \frac{ e V_\mathrm{S,2}}{k_\mathrm{B} T_\mathrm{e}}\right)}, \\
  \frac{d I_\mathrm{pl}}{d t} &=& L_\mathrm{pl}^{-1}(V_\mathrm{pl} + V_\mathrm{S,1} - V_\mathrm{S,2}) - \nu_\mathrm{eff} I_\mathrm{pl},
\end{eqnarray}
with $j_\mathrm{i0} = e n u_\mathrm{B} = e n \left(k_\mathrm{B} T_e/m_\mathrm{i} \right)^\frac{1}{2}$ and $j_\mathrm{e0}=e n \bar{v}_\mathrm{e}$ assuming a Maxwellian electron energy distribution function. $u_\mathrm{B}$ is the Bohmvelocity and $m_\mathrm{Ar}$ the argon ion mass. The sheath capacitances $C_\mathrm{S,1} = \left( 2 e n \varepsilon_0 A_\mathrm{E}^2/V_\mathrm{S,1} \right)^\frac{1}{2}$ and $C_\mathrm{S,2} = \left( 2 e n \varepsilon_0 A_\mathrm{G}^2 / V_\mathrm{S,2} \right)^\frac{1}{2}$ depend on the sheath voltages $V_\mathrm{S,1}$ and $V_\mathrm{S,2}$ as well as the electrode areas seen by the plasma in front of the driven electrode, $A_\mathrm{E}$, and the ground, $A_\mathrm{G}$. The inductance $L_\mathrm{pl}= l_\mathrm{B} m_\mathrm{e} / \mathrm{e}^2 n A_\mathrm{E}$ and the resistance $R_\mathrm{pl} = \nu_\mathrm{eff} L_\mathrm{pl}$ depend both on the bulk length $l_\mathrm{B}$ and on $A_\mathrm{E}$, which is due to the assumption of a homogeneous, cylindrical discharge. $I_\mathrm{pl}$ is the accumulated current flowing through the whole discharge and $V_\mathrm{pl}$ the voltage at the driven electrode.

\begin{widetext}
\begin{figure}[t!]
	\centering
	\resizebox{16cm}{!}{		
		\includegraphics[width=8cm]{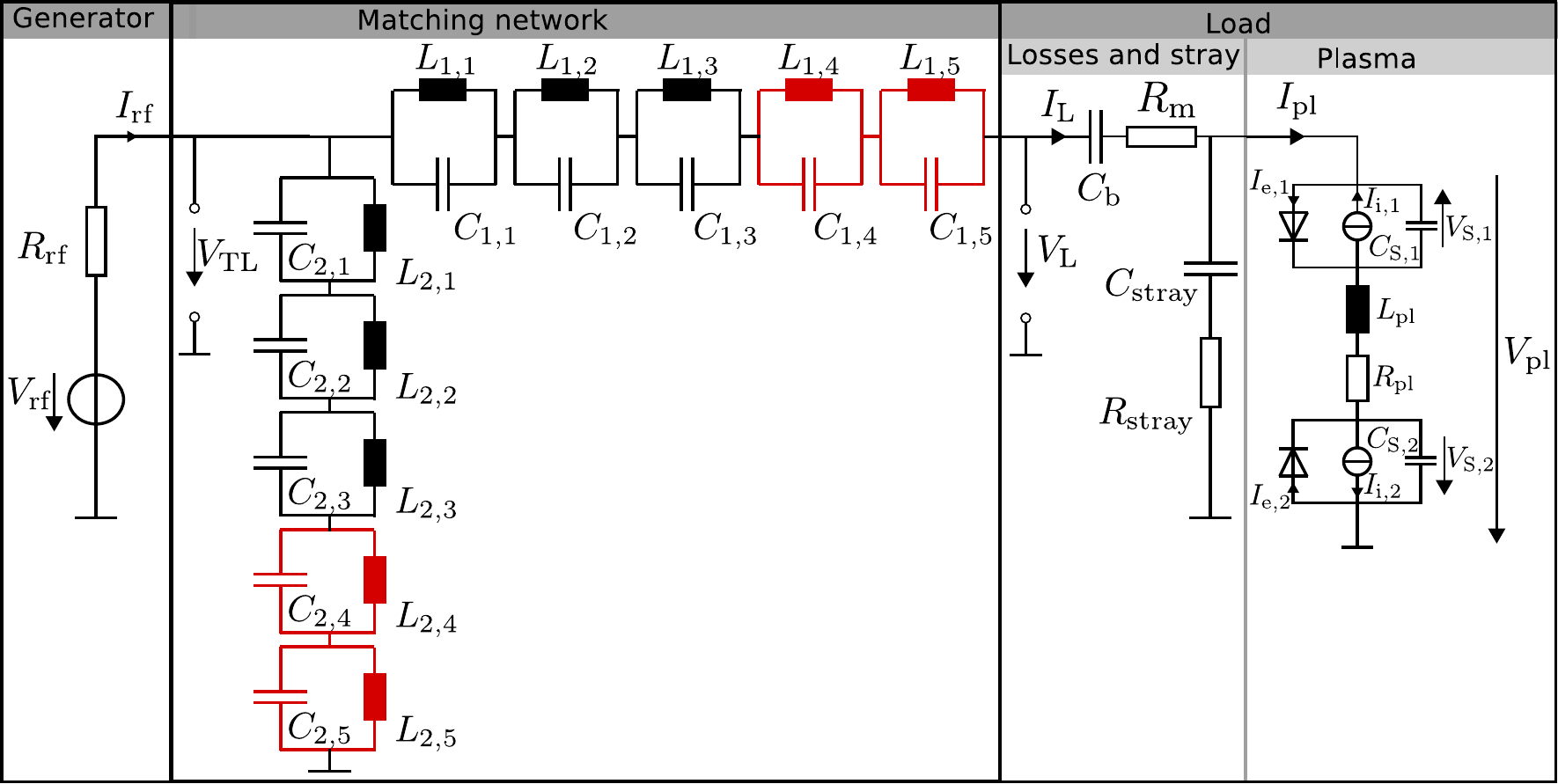}
	}
	\caption{Complete network that is simulated. The elements in red are only included for the simulation case with five excitation frequencies. The equivalent circuit of the plasma is on the right hand side.}
	\label{fig:setup}
\end{figure}
\end{widetext}

The equivalent circuit for this plasma model is depicted in figure~\ref{fig:setup} on the very right hand side. Notice that in this plot the scalar currents are defined by $I_\mathrm{e,1}=j_\mathrm{e0} A_\mathrm{E} \exp{\left(-e V_\mathrm{S,1}/k_\mathrm{B} T_\mathrm{e}\right)}$, $I_\mathrm{e,2}=j_\mathrm{e0} A_\mathrm{G} \exp{\left(-e V_\mathrm{S,2}/k_\mathrm{B} T_\mathrm{e}\right)}$, $I_\mathrm{i,1}=j_\mathrm{i0} A_\mathrm{E}$, and $I_\mathrm{i,2}=j_\mathrm{i0} A_\mathrm{G}$.
All elements of the model are functions of the electron temperature and electron density, which in turn depend on the external parameters of the discharge and the absorbed power. The latter is of special interest in this work, because it is a measure of the quality of the matching. Therefore, the simulation model is expanded in order to catch these dependencies.
The absorbed power $P_\mathrm{abs}$ of the plasma can be calculated by~\cite{lieberman_principles_2005}
\begin{align}
	P_\mathrm{abs} = n V_\mathrm{p} n_\mathrm{Ar} K_\mathrm{iz} \mathcal{E}_\mathrm{T},
\label{eq:power_balance}
\end{align}
with the plasma volume $V_\mathrm{p}$, the argon neutral gas density $n_\mathrm{Ar}$, the total energy lost in the system  per electron-ion pair created $\mathcal{E}_\mathrm{T}=\mathcal{E}_\mathrm{T}(T_\mathrm{e}, V_\mathrm{S,1}, V_\mathrm{S,2})$ and the ionization rate $K_\mathrm{iz}(T_\mathrm{e})$. A detailed analysis of $\mathcal{E}_\mathrm{T}$ in this setup can be found elsewhere~\cite{schmidt_consistent_2018}. Equating the particles created to the particles lost in the system, an equation
\begin{align}
	V_\mathrm{p} n_\mathrm{Ar} K_\mathrm{iz}= u_\mathrm{B} A 
\label{eq:particle_balance}
\end{align}
can be found, which accounts for particle conservation, with the total area around the discharge $A=A_\mathrm{E}+A_\mathrm{G}$.

\begin{figure}[t!]
	\centering
	\resizebox{12cm}{!}{		
		\includegraphics[width=8cm]{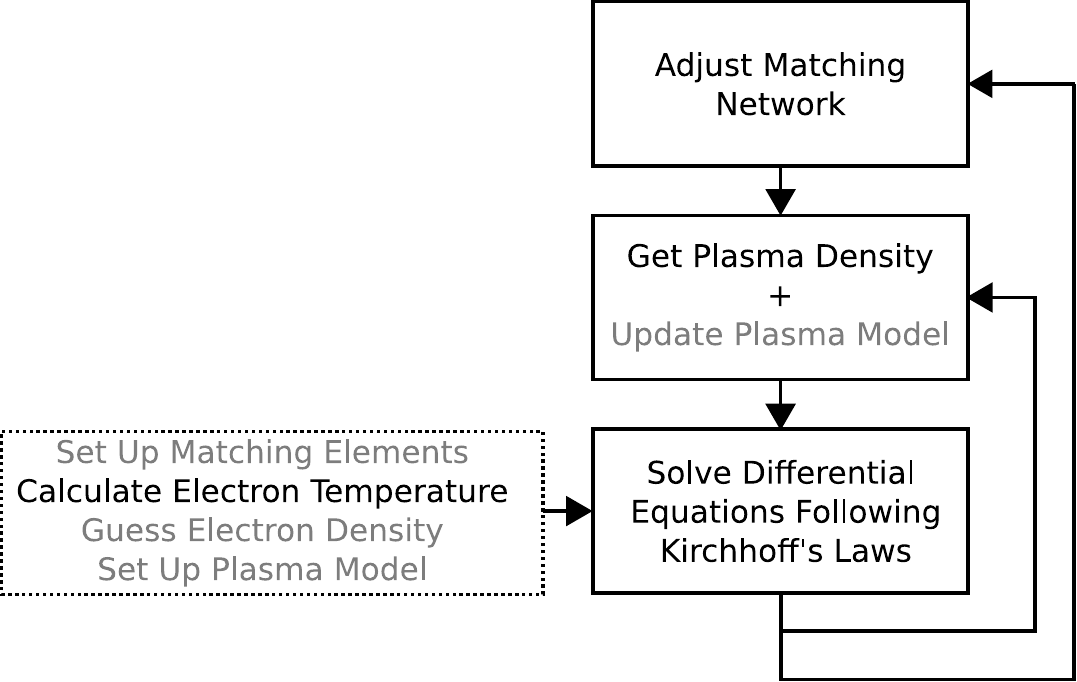}
	}
	\caption{Flowchart of the algorithm for the simulation. The steps in black indicate calculations performed within the simulation algorithm. The steps in grey represent parameter adjustments.}
	\label{fig:algorithm}
\end{figure}

The simulation now must include equations~\eqref{eq:power_balance} and \eqref{eq:particle_balance} in the circuit simulation as well as the elements of the matching network. This is achieved iteratively in different loops. The complete algorithm of the simulation is depicted in figure~\ref{fig:algorithm}. First, the initial values of all elements must be defined. Therefore, both $n$ and $T_\mathrm{e}$ need to be specified. If the geometrical parameters $A_\mathrm{E}$, $A_\mathrm{G}$ and $V_\mathrm{pl}$ are determined based on a given reactor geometry, the electron temperature $T_\mathrm{e}$ can be calculated from equation~\eqref{eq:particle_balance}, as $K_\mathrm{iz}(T_\mathrm{e})$ and $u_\mathrm{B}(T_\mathrm{e})$ are functions of the electron temperature. With the assumption that the geometrical parameters do not vary with the absorbed power, i.e., the plasma volume and surface area stay the same, equation~\eqref{eq:particle_balance} only needs to be evaluated once to specify $T_\mathrm{e}$. The plasma density $n$ is obtained from equation~\eqref{eq:power_balance} for the given set of parameters. The  matching elements and plasma density initially need to be assigned arbitrary (reasonable) values. The goal of the two loops of the simulation algorithm is to iteratively find the correct values for $n$ and the matching elements, respectively.
This is achieved by first simulating the whole circuit, whose elements all have defined values at this point, solving Kirchhoff's differential equations for the whole network in time-domain making use of the software ngSPICE~\cite{vogt_ngspice_2017}. The simulation provides the current and voltage at every point in the system. So the absorbed power in the plasma can be calculated by $P_\mathrm{abs} = \int dt  I_\mathrm{pl} V_\mathrm{pl}$, with $I_\mathrm{pl}$ being the current flowing through the plasma model and $V_\mathrm{pl}$ the respective voltage. Using this value for $P_\mathrm{abs}$, equation~\eqref{eq:power_balance} is then used to calculate the plasma density $n$ and update the elements of the plasma model. This step is repeated until convergence is reached, i.e., the plasma density and the absorbed power do not change anymore. This is the inner circle of the simulation algorithm. In the outer loop a fast Fourier transform (FFT) of the currents and voltages in the system can be used to calculate the complex impedance of the load for every excitation frequency in the form of $Z_{\mathrm{L},i} =V_{\mathrm{L},i}/I_{\mathrm{L},i}$, with the current $I_{\mathrm{L},i}$ flowing through the load and the voltage $V_{\mathrm{L},i}$ across the load. Then, using equations~\eqref{eq:real_part} and \eqref{eq:im_part} and additional constraints (e.g., constant inductances), the elements of the matching network can be calculated and updated. Because this influences the power transferred into the plasma load, whose elements depend on the absorbed power, this step must be repeated until the results reach steady state.

\section{\label{sec:setup} Exemplary Setups and Simulation Results}


\begin{table}
\begin{tabular}{l | c}
\toprule
Parameter & Value \\
\colrule
$T_\mathrm{Ar}$ & 300 K \\
$A_\mathrm{E}$ & 707 cm$^2$\\
$A_\mathrm{G}$ & 1650 cm$^2$ \\
$l_\mathrm{B}$ & 10 cm \\
$V_1$ & 300 V \\
$V_2$ & 300 V \\
$V_3$ & 300 V \\
$V_4$ & 300 V \\
$V_5$ & 300 V \\
$\omega_1$ & $2 \pi \times 13.56$~MHz \\
$\omega_2$ & $2 \pi \times 27.12$~MHz \\
$\omega_3$ & $2 \pi \times 40.68$~MHz \\
$\omega_4$ & $2 \pi \times 54.24$~MHz \\
$\omega_5$ & $2 \pi \times 67.80$~MHz \\
$R_\mathrm{rf}$ & $50~\Omega$ \\
$C_\mathrm{b}$ & $1000~$pF \\
$R_\mathrm{m}$ & $1~\Omega$ \\
$R_\mathrm{stray}$ & $0.5~\Omega$ \\
$C_\mathrm{stray}$ & $200$ pF \\
\botrule
\end{tabular}
\caption{Input parameters of the simulation}
\label{table:parameter}
\end{table}

In order to demonstrate the validity of the presented method, three different scenarios will be analyzed in this section: A) $N=3$ frequency excitation and pressure $p=200$~Pa, B) $N=3$ frequency excitation and pressure $p=1$~Pa, and C) $N=5$ frequency excitation and pressure $p=200$~Pa. The other remaining discharge and network parameters, which can be found in table~\ref{table:parameter}, are the same for all cases. Notice that $V_4=V_5=0$~V for the first two cases and the frequencies are harmonics of the fundamental frequency. All phase shifts for the respective excitations are set to  $\varphi_i = 0$.

The general design of the matching network follows the considerations of section~\ref{sec:method} for 3 frequencies and 5 frequencies respectively. For all three cases, $Z_1$ and $Z_2$ are composed of parallel resonance circuits, i.e., $M=N$, $k_{M+1}=0$ and $k_{0}=0$. Finally, the last case is used to study the difference between choosing this network and $M=N-1$ as well as $k_{M+1}$ and $k_{0}$ nonzero. Resistive losses in the network are represented by a  resistance $R_\mathrm{m}$. A resistance $R_\mathrm{stray}$ and a capacitance $C_\mathrm{stray}$ represent the influence of the reactor chamber. All of theses values are rough estimates and depend strongly on the actual reactor setup and elements in the network. Additionally, a capacitance $C_\mathrm{b}$ is included to allow a DC self-bias voltage to be built up. The exact value of this element does not matter as long as it is reasonably large. The whole network including all elements is shown in figure~\ref{fig:setup}. Notice that in this figure $M=N$. The elements $C_{2,4}$, $L_{2,4}$, $C_{2,5}$, $L_{2,5}$, $C_{1,4}$, $L_{1,4}$, $C_{1,5}$ and $L_{1,5}$ are only included for the last case with $N=5$.

\begin{figure}[t!]
	\centering
	\resizebox{12cm}{!}{		
		\includegraphics[width=12cm]{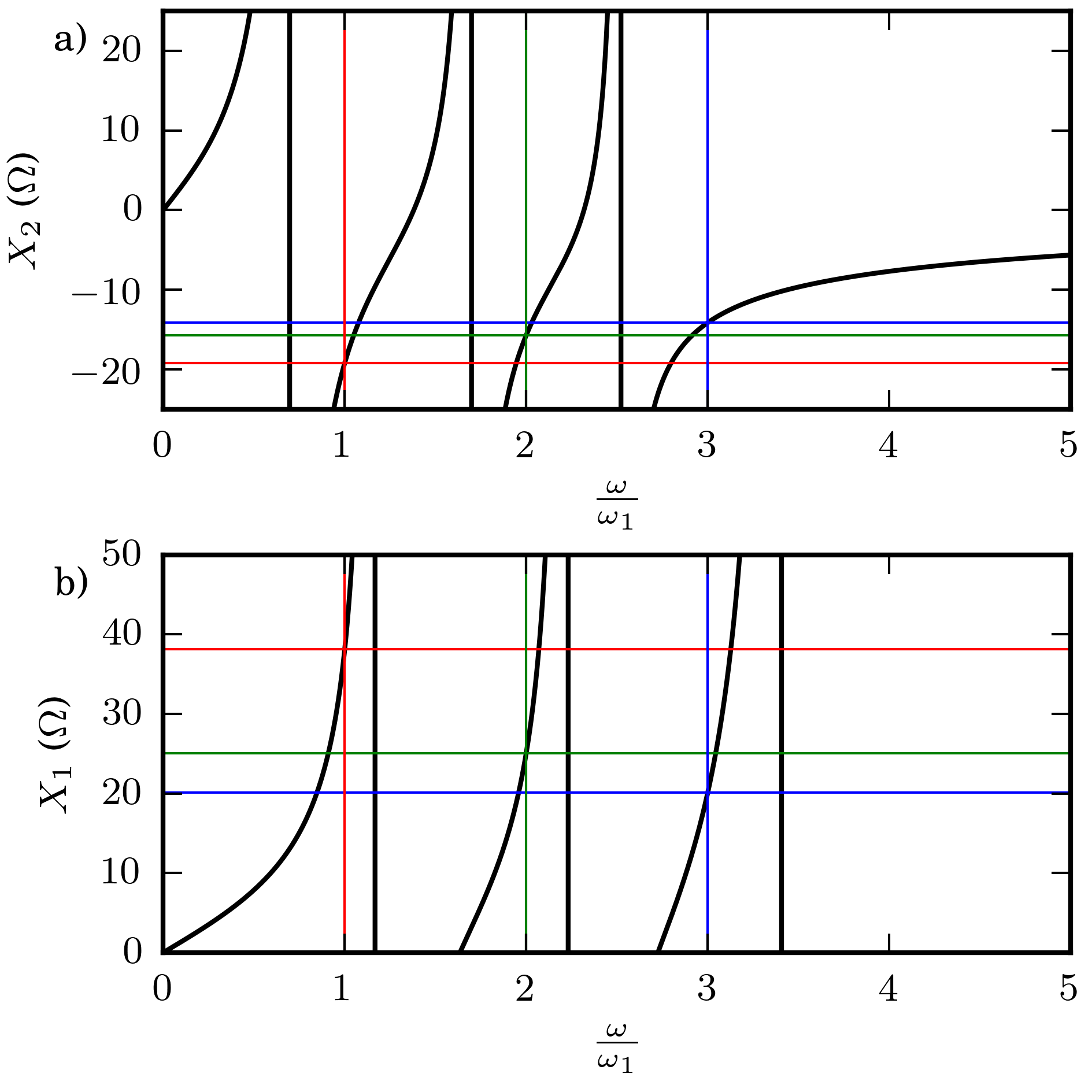}
	}
	\caption{Reactance-functions for $X_1$ and $X_2$ for a converged simulation of case A ($N=3$; $p=200$~Pa). At the different excitation frequencies the function has the values following equations~\eqref{eq:real_part} and \eqref{eq:im_part}, which is indicated by the horizontal lines.}
	\label{fig:impedances}
\end{figure}

\subsection{\label{sec:case1} Three frequencies, \boldsymbol{$p=200$}~Pa}

\begin{figure}[t!]
	\centering
	\resizebox{12cm}{!}{		
		\includegraphics[width=8cm]{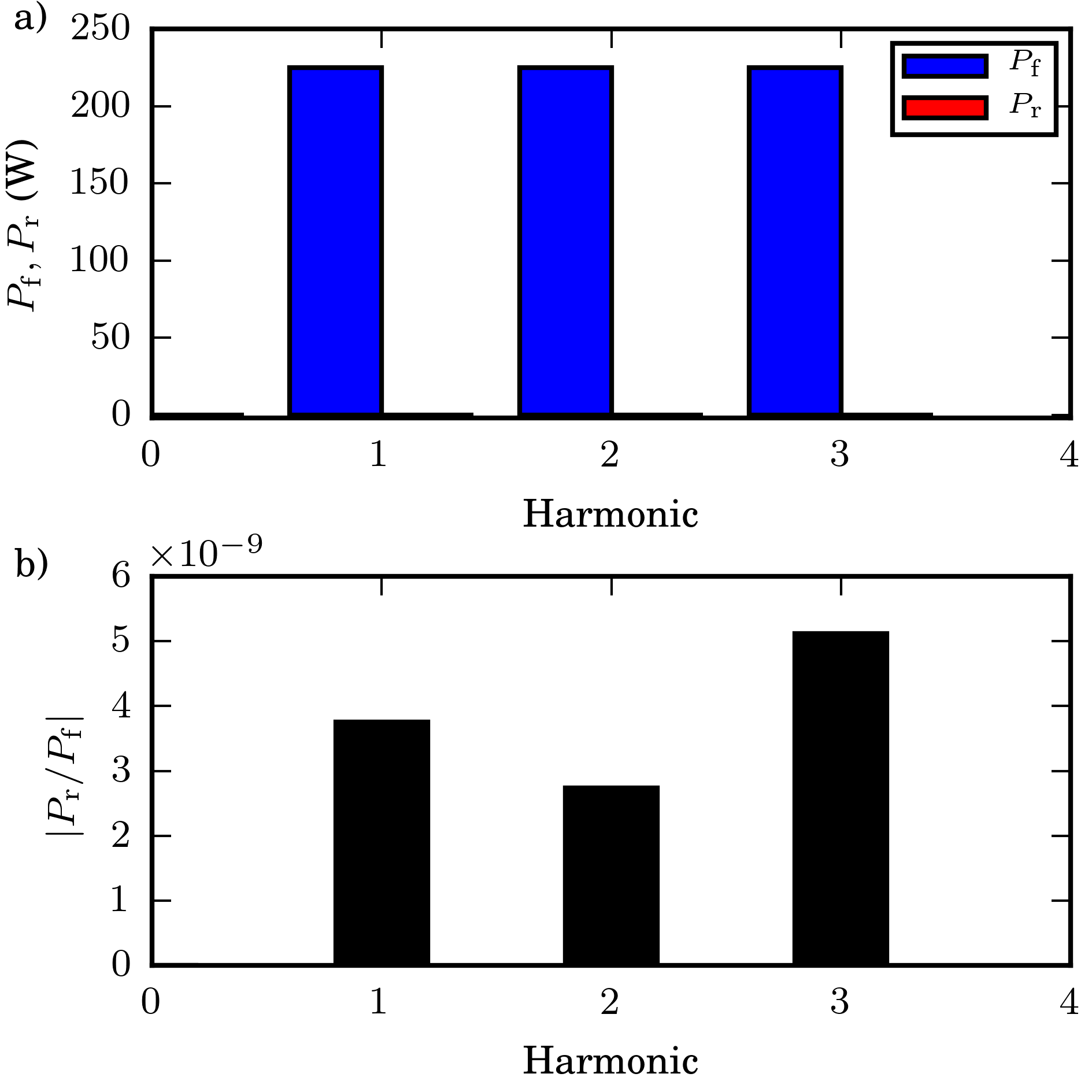}
	}
	\caption{Forward power $P_\mathrm{f}$, reflected power $P_\mathrm{r}$ and their ratios for three excitation frequencies and a pressure of $p=200$~Pa. $P_\mathrm{r}$ is small compared to $P_\mathrm{f}$, that is why it is hard to see in plot a).}
	\label{fig:power_high_p}
\end{figure}

Using the parameters from table~\ref{table:parameter} and the pressure $p=200$~Pa, equation~\eqref{eq:particle_balance} provides a value of $T_\mathrm{e} =1.8$~eV. For each of the $N=3$ frequency components, the FFT gives the load impedance consisting of a resistance and reactance. For a converged simulation these values amount to $R_{\mathrm{L},1}=6.43~\Omega$, $R_{\mathrm{L},2}=4.51~\Omega$, $R_{\mathrm{L},3}=3.71~\Omega$, $X_{\mathrm{L},1}=-21.36~\Omega$, $X_{\mathrm{L},2}=-10.74~\Omega$, and $X_{\mathrm{L},3}=-6.98~\Omega$. The plasma density for this case is $n=3.15 \times 10^{16}~\mathrm{m}^{-3}$.
Using these values, the left hand sides of equations~\eqref{eq:real_part} and \eqref{eq:im_part} result in $X_{2,1} = - 19.21~\Omega$, $X_{2,2} = - 15.73~\Omega$, $X_{2,3} = - 14.15~\Omega$, $X_{1,1} = 38.10~\Omega$, $X_{1,2} = - 25.06~\Omega$, and $X_{1,3} = - 20.08~\Omega$. For both $X_1$ and $X_2$ 
equation~\eqref{eq:transformation} with $k_0= 0$, $K_{M+1}=0$ and $M=N=3$ defines the respective network giving $N=3$ equations for each $X_1$ and $X_2$. Additional constraints are set to $L_{2,1} = 300$~nH, $L_{2,2} = 30$~nH, $L_{2,3} = 10$~nH, $L_{1,1}=100$~nH, and $L_{1,2}=L_{1,3}=30$~nH. In this way the two systems of equations can be solved for the six unknown capacitances, which result in $C_{2,1}=945$~pF, $C_{2,2}=1590$~pF, $C_{2,3}=2166$~pF, $C_{1,1}=1010$~pF, $C_{1,2}=923$~pF, and $C_{1,3}=395$~pF. The frequency-dependent reactance functions for $X_1$ and $X_2$ are depicted in figure \ref{fig:impedances}~a) and \ref{fig:impedances}~b), respectively. The functions have the values calculated above at each excitation frequency, and thereby define a network which satisfies the matching conditions. For a suitable practical application, different constraints, i.e., values of the inductances, may be more reasonable. The proposed exemplary values may be realized with simple transmission lines without the need of actual (bulk) elements, with the advantage of being less lossy. Having said that, other issues in practice might be of relevance that are not included in this reasoning. However, the choice of the inductance values is not completely free, which will be considered in detail in the discussion of simulation case \textbf{C}.

Setting up the elements of the matching network in the described manner, the impedance of the total network can be calculated to $Z_{\mathrm{TL},i} = I_{\mathrm{rf},i}/V_{\mathrm{TL},i}$, which is the total load impedance including the matching network. Again, for ideal matching the impedance at each excitation frequency should be equal to $R_\mathrm{rf} = 50~\Omega$. Using the simulation results, the impedances for each component are be calculated to be $Z_\mathrm{TL,1} = (50.006 + j 0.0006)~\Omega$, $Z_\mathrm{TL,2} = (50.003- j 0.004)~\Omega$, and $Z_\mathrm{TL,3} = (50.003 - j 0.007)~\Omega$. All are reasonably close to $R_\mathrm{rf} = 50~\Omega$.

Also an analysis of the forward power $P_\mathrm{f}$ consumed in the total load (including the matching network) and the reflected power $P_\mathrm{r}$ offers additional insight. For ideal matching the reflected power should  be $P_\mathrm{r}=0$. Both $P_\mathrm{f}$ and $P_\mathrm{r}$ as well as their ratio are depicted in figure~\ref{fig:power_high_p}. $P_\mathrm{f} \approx 225$~W for all three frequency components, while $|P_\mathrm{r}/P_\mathrm{f}|< 10^{-8}$, which is very small, showing that the matching is nearly ideal. It should be noted that in reality such a value will be hard to achieve, due to various effects that are not included in the simulation such as elements, which are not ideal and not strictly linear. The value of $225~$W is expected for an ideally matched load, since in this case $P_\mathrm{f} \approx V_{\mathrm{TL},i}^2/2|Z_{\mathrm{TL},i}| = 225~$W for $V_{\mathrm{TL},i} = V_i/2$ and $|Z_{\mathrm{TL},i}|$ assumed to be $50~\Omega$.
The observed deviations in the power and the total load impedance from the ideal case are mainly due to numerical inaccuracies. This could be improved by increasing the simulation time and reducing the time step and would lead to better averaging in the FFT and higher accuracy in the circuit simulation.

\subsection{\label{sec:case2} Three frequencies, \boldsymbol{$p=1$}~Pa}

\begin{figure}[t!]
	\centering
	\resizebox{12cm}{!}{		
		\includegraphics[width=8cm]{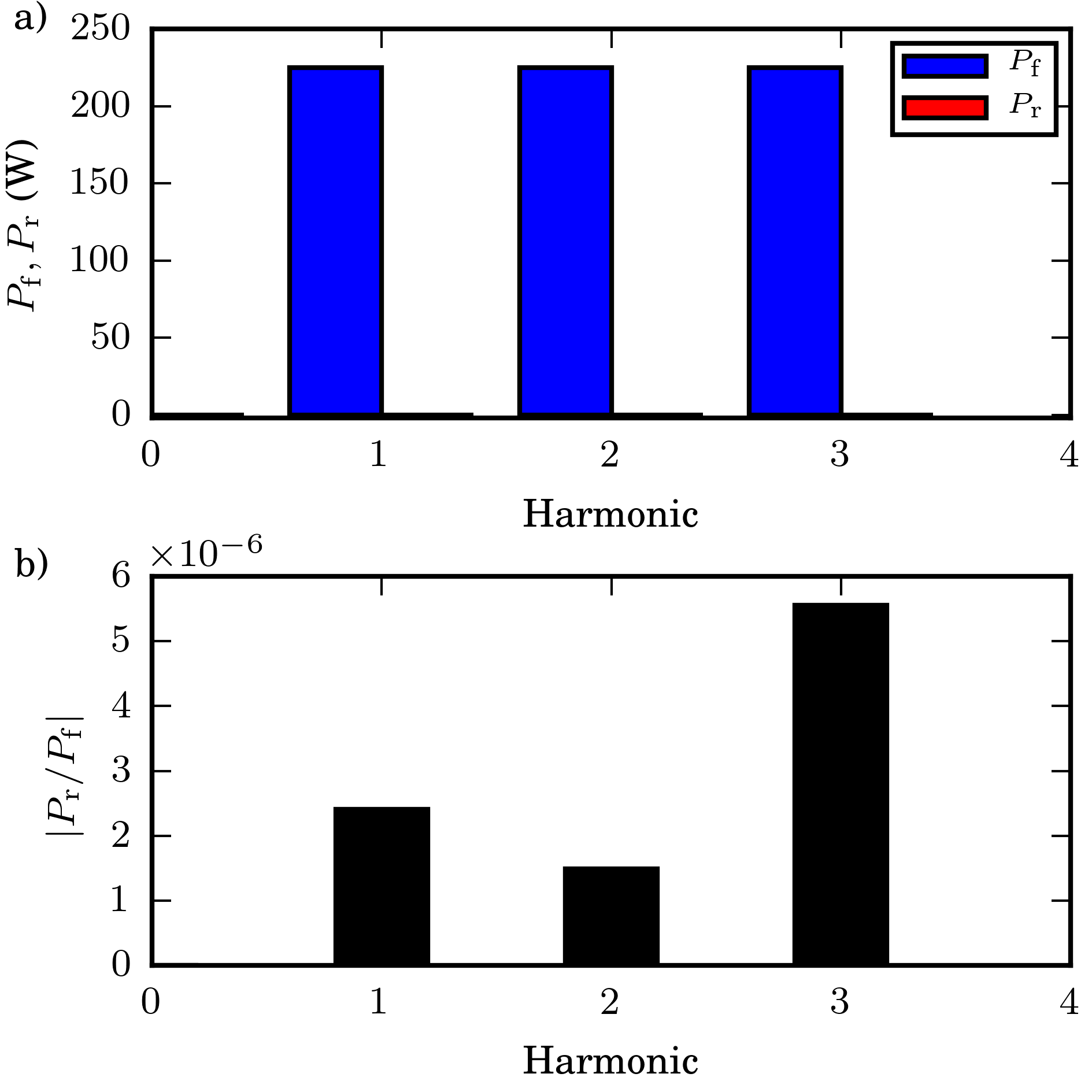}
	}
	\caption{Forward power $P_\mathrm{f}$, reflected power $P_\mathrm{r}$ and their ratios for three excitation frequencies and a pressure of $p=1$~Pa. $P_\mathrm{r}$ is small compared to $P_\mathrm{f}$, that is why it is hard to see in plot a).}
	\label{fig:power_low_p}
\end{figure}

In this case the pressure of $p=1$~Pa leads to a plasma with stronger nonlinearities and, consequently, significant plasma-generated harmonics in the plasma current. This is mainly due to the plasma series resonance, which is only of significance at low pressures whereas it is damped by collisions at higher pressures. The simulation of this case is therefore interesting in order to see if the proposed matching method also works under these nonlinear conditions. All other plasma parameters are the same as in case A).

The electron temperature is calculated to be $T_\mathrm{e}=3.6$~eV. A converged simulation leads to a plasma density of $n=1.03 \times 10^{16}~\mathrm{m}^{-3}$ and a load with $R_{\mathrm{L},1}=4.90~\Omega$, $R_{\mathrm{L},2}=2.16~\Omega$, and $R_{\mathrm{L},3}=0.72~\Omega$, $X_{\mathrm{L},1}=-34.58~\Omega$, $X_{\mathrm{L},2}=-16.89~\Omega$, and $X_{\mathrm{L},3}=-10.70~\Omega$. Again, the system of equations~\eqref{eq:real_part} and \eqref{eq:im_part} is used to calculate the impedance functions for the matching network. The chosen constraints for this case are $L_{2,1}= 100$~nH, $L_{2,2}= 20$~nH, $L_{2,3} = 10$~nH, $L_{1,1}=100$~nH, $L_{1,2}= 30$~nH and $L_{1,3}=30$~nH. The values of the capacitances can then be calculated to $C_{2,1}=2270$~pF, $C_{2,2}=2040$~pF, $C_{2,3}=2990$~pF, $C_{1,1}=625$~pF, $C_{1,2}=938$~pF, and $C_{1,3}=375$~pF. 

Using these values, the impedance of the total load amounts to $Z_\mathrm{TL,1} = (50.09 - j 0.12)~\Omega$, $Z_\mathrm{TL,2} = (50.12 - j 0.007)~\Omega$ and $Z_\mathrm{TL,3} = (50.01 - j 0.23)~\Omega$, which is again reasonably close to $R_\mathrm{rf} = 50~\Omega$. The forward and reflected powers $P_\mathrm{f}$ and $P_\mathrm{r}$ are presented in figure~\ref{fig:power_low_p}. $P_\mathrm{f} \approx 225$~W and $|P_\mathrm{r}/P_\mathrm{f}|<10^{-5}$ is maintained for all frequencies. Again, the deviations from the ideal result can be explained by numerical inaccuracies. Also the fact that the load is very nonlinear compared to the other cases may explain some of the deviations: The plasma is not only a passive load, but also a generator. While it still absorbs more power than it generates, the fact that it is not strictly passive may cause some issues, since an impedance is by definition a passive element.

The plasma also generates harmonics greater than $3 \, \omega_1$. Most of the power that these harmonics dissipate remains within the load, i.e., in the loss elements, and very little actually reaches the generator. Therefore, it does not cause any issues regarding the matching. These effects will be analyzed in more detail in a future publication.

\subsection{\label{sec:case3} Five frequencies, \boldsymbol{$p=200$}~Pa}

\begin{figure}[t!]
	\centering
	\resizebox{12cm}{!}{		
		\includegraphics[width=8cm]{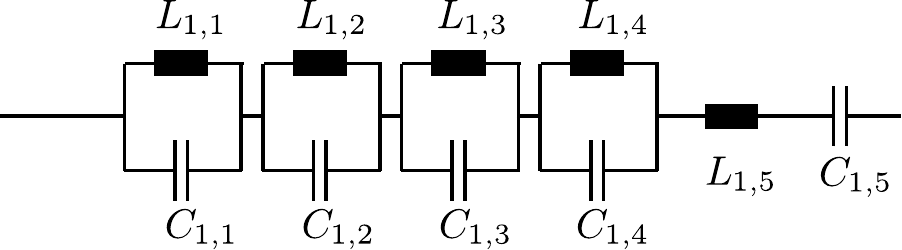}
	}
	\caption{Alternative design of the matching branch $Z_1$ displayed in figure~\ref{fig:setup}.}
	\label{fig:network2}
\end{figure}

\begin{figure}[t!]
	\centering
	\resizebox{12cm}{!}{		
		\includegraphics[width=8cm]{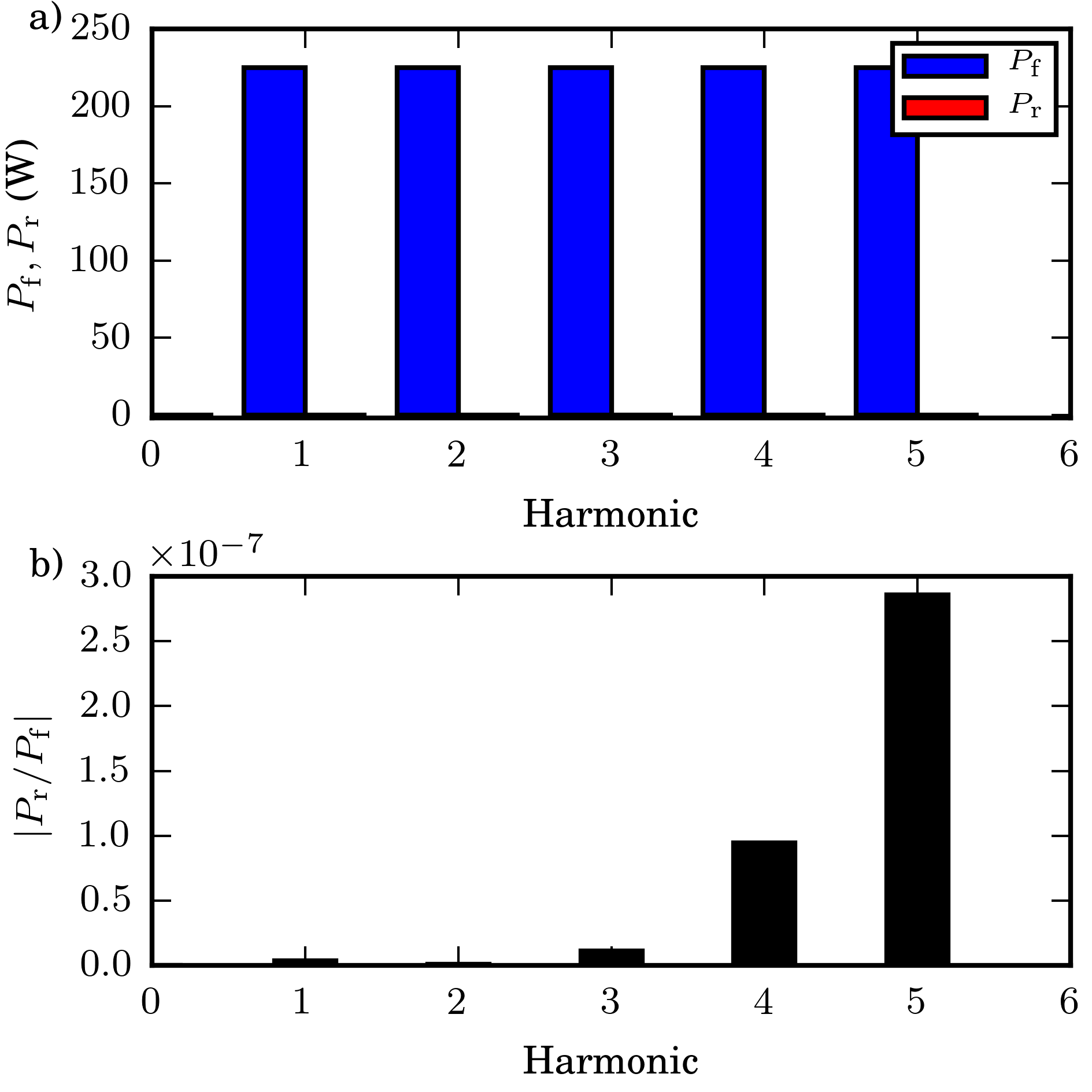}
	}
	\caption{Forward power $P_\mathrm{f}$, reflected power $P_\mathrm{r}$ and their ratios for five excitation frequencies and a pressure of $p=200$~Pa. $P_\mathrm{r}$ is small compared to $P_\mathrm{f}$, that is why it is hard to see in plot a).}
	\label{fig:power_high_p_n5}
\end{figure}

The higher the number of excitation frequencies, the more accurately an idealized voltage waveform can be approximated. Therefore, it is instructive to investigate the case of $N=5$ harmonics. The pressure in this exemplary case is set to $p=200$~Pa, which gives an electron temperature of $T_\mathrm{e}=1.8$~eV.
For a converged simulation the load amounts to $n=4.81 \times 10^{16}~\mathrm{m}^{-3}$, $R_{\mathrm{L},1}=5.51~\Omega$, $R_{\mathrm{L},2}=3.83~\Omega$, $R_{\mathrm{L},3}=3.33~\Omega$, $R_{\mathrm{L},4}=3.09~\Omega$, $R_{\mathrm{L},5}=2.85~\Omega$, $X_{\mathrm{L},1}=-19.61~\Omega$, $X_{\mathrm{L},2}=-9.87~\Omega$, $X_{\mathrm{L},3}=-6.66~\Omega$, $X_{\mathrm{L},4}=-4.91~\Omega$, and $X_{\mathrm{L},5}=-3.77~\Omega$. Systems of equations~\eqref{eq:real_part} and \eqref{eq:im_part} can be solved for $N=5$ using the constraints $L_{2,1}= 60$~nH, $L_{2,2}= 30$~nH, $L_{2,3} = 20$~nH, $L_{2,4} = 20$~nH, $L_{2,5} = 20$~nH, $L_{1,1}=13$~nH, $L_{1,2}= 7$~nH, $L_{1,3}=4$~nH, $L_{1,4}= 2$~nH and $L_{1,5}=13$~nH. While these values can be simulated and lead to plausible results, it is in practice hard to build a network using inductances that small. However, the use of larger values for the inductances $L_{1,2}$, $L_{1,3}$ and $L_{1,4}$ leads to a system of equations, which cannot be consistently solved. In order to understand the reason for this, it is instructive to maintain $Z_2$, but to change the design of the $Z_1$ network by replacing one LC-parallel unit with an LC-series unit and compare the behavior of this new branch with the previous one. In terms of the definition following from equation~\eqref{eq:transformation}, $M=N-1$ is chosen and two additional series elements are used for $Z_1$, that is $k_0$ and $k_\mathrm{M+1}$ are both nonzero. This leads to the network shown in figure~\ref{fig:network2}. Solving the systems of equations using this network, the constraints are satisfied with notably larger values $L_{1,1}=400$~nH, $L_{1,2}= 150$~nH, $L_{1,3}=60$~nH, $L_{1,4}= 30$~nH and $L_{1,5}=300$~nH. To some degree $L_{1,5}$ can be used to influence the value of the other inductances: If $L_{1,5}$ is increased, the other inductances can be increased as well. This effect can be understood by interpreting $L_{1,5}$ not as a part of the matching network, but instead as part of the load. Therefore the ``load'' reactance is increased by $\omega L_{1,5}$. The remaining elements of network $Z_1$ are displaced according to this additional load. Since the positions of the poles defined by $q_m =1/C_m L_m$ must be within certain bounds, e.g., in between the excitation frequencies, $Z_1$ is predominantly influenced as required by the respective $k_m = 1/C_m$. If the absolute value of $Z_1$ needs to be increased at one or more frequencies, $k_m$ must increase resulting in a decrease in $C_m$. As a result, $L_m$ must increase in order for the poles to not change their positions. Since $\omega L_{1,5}$ can be used to determine how large the ``amplitude'' of the resonance circuits needs to be, it can control the size of the other elements in the $Z_1$ branch.

This also explains the different values of the respective inductances in the different cases: By changing $C_m$, not only the ``amplitude'' of a specific resonance is changed, but also the position of the poles. Therefore, for constant $L_m$ only a limited parameter range can be reached by only changing the capacitances. If one desires to build a matching network in practice, some assumptions regarding the load impedances are helpful in order to choose meaningful values for the inductances. A specific matching network could then be equipped with easy-to-change inductances in order to make it applicable for different scenarios. 

The specific choice for the network, either using only parallel resonance units or an additional series resonance circuit, depends on the reactor characteristics and on the desired elements used to build the network. A large inductance might be easier to build, but is also lossier, which might in some applications be of high relevance. It is therefore important to keep in mind that the values chosen in this work are mainly demonstrative and in practice need to be adjusted accordingly within the general framework. The observation that the case with five frequency components required small inductances for a specific design of $Z_1$ depends strongly on the details of the specific plasma chamber, the plasma parameters and the loss elements. In another scenario with different plasma and chamber properties this might not occur. Also this specific problem might be avoided by choosing a completely different method for synthesizing $Z_1$ and $Z_2$. 

Choosing the design for $Z_1$ depicted in figure~\ref{fig:network2} and the above listed constraints, the capacitances in the present scenario amount to: $C_{2,1}=940$~pF, $C_{2,2}=780$~pF, $C_{2,3}=486$~pF, $C_{2,4}=378$~pF, $C_{2,5}=175$~pF, $C_{1,1}=144$~pF, $C_{1,2}=152$~pF, $C_{1,3}=180$~pF, $C_{1,4}=231$~pF and $C_{1,5}=112$~pF. Notice, that for a branch $Z_1$ with a series resonance circuit, the capacitances are considerably smaller, which might also be useful in practice.
The power distribution for the different excitation frequencies is shown in figure~\ref{fig:power_high_p_n5}. The forward power is $P_\mathrm{f} \approx 225$~W for all five frequencies, while $|P_\mathrm{r}/P_\mathrm{f}| <10^{-6} $. As can be observed, the value of $|P_\mathrm{r}/P_\mathrm{f}| $ increases with the frequency. This trend is probably due to the internal calculations in ngSPICE and in the FFT algorithm being more inaccurate for higher frequencies, due to the constant time step of the simulation being bigger in relation for larger frequencies.

All these results prove that effective matching conditions can be achieved using the proposed method for the design of a multiple frequency matching network.  

\section{\label{sec:conclusion} Conclusion}

A method is proposed for coupling a multiple frequency excitation to a frequency dependent complex load, which is especially suited for voltage waveform tailoring in radio-frequency plasma sources. In order to simulate such a network attached to a plasma source, a simulation method is developed based on an equivalent plasma circuit model extended by a self-consistent calculation of the electron temperature and plasma density. The algorithm of this simulation method is discussed, including  loss and stray effects. Three different scenarios are then simulated to prove the viability and feasibility of the proposed method. In all cases, the matching conditions can be reached to a very high accuracy, which proves the functionality of the matching method. This allows voltage waveform tailoring to be used in industrial applications via an arbitrary function generator and a broadband amplifier.

\section*{\label{sec:acknowledgments} Acknowledgments}
This work was supported and funded by Lam Research, the DFG (German Research Foundation) within the framework of the Collaborative Research Centres SFB 1316 (projects A4 and A5) and TRR 87 (project C1 and C8) as well as by the US National Science Foundation (grant PHY 1601080). The authors thank R. P. Brinkmann and T. Gergs from the Institute of Theoretical Electrical Engineering, Ruhr University Bochum for fruitful discussions.

\section*{ORCID IDs}
\href{http://orcid.org/0000-0001-6445-4990}{T. Mussenbrock: http://orcid.org/0000-0001-6445-4990}\\
\href{https://orcid.org/0000-0001-9136-8019}{J. Trieschmann: https://orcid.org/0000-0001-9136-8019}

\bibliographystyle{aip}

\bibliography{Multi_f_matching.bib}

\begin{thebibliography}{10}

\bibitem{lieberman_principles_2005}
M.~A. Lieberman and A.~J. Lichtenberg,
\newblock {\em Principles of plasma discharges and materials processing},
\newblock Wiley-Interscience, Hoboken, N.J, 2nd ed edition, 2005.

\bibitem{chabert_physics_2011}
P.~Chabert and N.~Braithwaite,
\newblock {\em Physics of radio-frequency plasmas},
\newblock Cambridge University Press, Cambridge, 2011.

\bibitem{samukawa_2012_2012}
S.~Samukawa et~al.,
\newblock Journal of Physics D: Applied Physics {\bf 45}, 253001 (2012).

\bibitem{patterson_arbitrary_2007}
M.~M. Patterson, H.-Y. Chu, and A.~E. Wendt,
\newblock Plasma Sources Science and Technology {\bf 16}, 257 (2007).

\bibitem{wang_control_2000}
S.-B. Wang and A.~E. Wendt,
\newblock Journal of Applied Physics {\bf 88}, 643 (2000).

\bibitem{heil_numerical_2008}
B.~G. Heil, J.~Schulze, T.~Mussenbrock, R.~P. Brinkmann, and U.~Czarnetzki,
\newblock IEEE Transactions on Plasma Science {\bf 36}, 1404 (2008).

\bibitem{heil_possibility_2008}
B.~G. Heil, U.~Czarnetzki, R.~P. Brinkmann, and T.~Mussenbrock,
\newblock Journal of Physics D: Applied Physics {\bf 41}, 165202 (2008).

\bibitem{donko_pic_2009}
Z.~Donk{\'o}, J.~Schulze, B.~G. Heil, and U.~Czarnetzki,
\newblock Journal of Physics D: Applied Physics {\bf 42}, 025205 (2009).

\bibitem{schulze_electrical_2009}
J.~Schulze, E.~Sch{\"u}ngel, and U.~Czarnetzki,
\newblock Journal of Physics D: Applied Physics {\bf 42}, 092005 (2009).

\bibitem{schungel_electrical_2012}
E.~Sch{\"u}ngel, D.~Eremin, J.~Schulze, T.~Mussenbrock, and U.~Czarnetzki,
\newblock Journal of Applied Physics {\bf 112}, 053302 (2012).

\bibitem{schungel_power_2011}
E.~Sch{\"u}ngel, J.~Schulze, Z.~Donk{\'o}, and U.~Czarnetzki,
\newblock Physics of Plasmas {\bf 18}, 013503 (2011).

\bibitem{schungel_effect_2013}
E.~Sch{\"u}ngel, S.~Mohr, S.~Iwashita, J.~Schulze, and U.~Czarnetzki,
\newblock Journal of Physics D: Applied Physics {\bf 46}, 175205 (2013).

\bibitem{lafleur_tailored-waveform_2016}
T.~Lafleur,
\newblock Plasma Sources Science and Technology {\bf 25}, 013001 (2016).

\bibitem{schulze_electrical_2011}
J.~Schulze, E.~Sch{\"u}ngel, Z.~Donk{\'o}, and U.~Czarnetzki,
\newblock Plasma Sources Science and Technology {\bf 20}, 015017 (2011).

\bibitem{lafleur_separate_2012}
T.~Lafleur, P.~A. Delattre, E.~V. Johnson, and J.~P. Booth,
\newblock Applied Physics Letters {\bf 101}, 124104 (2012).

\bibitem{diomede_radio-frequency_2014}
P.~Diomede, D.~J. Economou, T.~Lafleur, J.-P. Booth, and S.~Longo,
\newblock Plasma Sources Science and Technology {\bf 23}, 065049 (2014).

\bibitem{berger_experimental_2015}
B.~Berger et~al.,
\newblock Journal of Applied Physics {\bf 118}, 223302 (2015).

\bibitem{derzsi_electron_2013}
A.~Derzsi, I.~Korolov, E.~Sch{\"u}ngel, Z.~Donk{\'o}, and J.~Schulze,
\newblock Plasma Sources Science and Technology {\bf 22}, 065009 (2013).

\bibitem{delattre_radio-frequency_2013}
P.-A. Delattre, T.~Lafleur, E.~Johnson, and J.-P. Booth,
\newblock Journal of Physics D: Applied Physics {\bf 46}, 235201 (2013).

\bibitem{bruneau_ion_2014}
B.~Bruneau, T.~Novikova, T.~Lafleur, J.~P. Booth, and E.~V. Johnson,
\newblock Plasma Sources Science and Technology {\bf 23}, 065010 (2014).

\bibitem{bruneau_control_2014}
B.~Bruneau, T.~Novikova, T.~Lafleur, J.~P. Booth, and E.~V. Johnson,
\newblock Plasma Sources Science and Technology {\bf 24}, 015021 (2014).

\bibitem{bruneau_strong_2015}
B.~Bruneau et~al.,
\newblock Physical Review Letters {\bf 114} (2015).

\bibitem{schungel_customized_2015}
E.~Sch{\"u}ngel et~al.,
\newblock Plasma Sources Science and Technology {\bf 24}, 045013 (2015).

\bibitem{schungel_prevention_2015}
E.~Sch{\"u}ngel, S.~Mohr, J.~Schulze, and U.~Czarnetzki,
\newblock Applied Physics Letters {\bf 106}, 054108 (2015).

\bibitem{iwashita_transport_2013}
S.~Iwashita et~al.,
\newblock Journal of Physics D: Applied Physics {\bf 46}, 245202 (2013).

\bibitem{iwashita_sheath--sheath_2012}
S.~Iwashita et~al.,
\newblock Plasma Sources Science and Technology {\bf 21}, 032001 (2012).

\bibitem{ahr_influence_2015}
P.~Ahr, E.~Sch{\"u}ngel, J.~Schulze, T.~V. Tsankov, and U.~Czarnetzki,
\newblock Plasma Sources Science and Technology {\bf 24}, 044006 (2015).

\bibitem{berger_enhanced_2017}
B.~Berger et~al.,
\newblock Applied Physics Letters {\bf 111}, 201601 (2017).

\bibitem{johnson_nanocrystalline_2010}
E.~V. Johnson, T.~Verbeke, J.-C. Vanel, and J.-P. Booth,
\newblock Journal of Physics D: Applied Physics {\bf 43}, 412001 (2010).

\bibitem{johnson_microcrystalline_2012}
E.~V. Johnson, P.~A. Delattre, and J.~P. Booth,
\newblock Applied Physics Letters {\bf 100}, 133504 (2012).

\bibitem{johnson_hydrogenated_2012}
E.~Johnson, S.~Pouliquen, P.~Delattre, and J.~Booth,
\newblock Journal of Non-Crystalline Solids {\bf 358}, 1974 (2012).

\bibitem{hrunski_influence_2013}
D.~Hrunski et~al.,
\newblock Thin Solid Films {\bf 532}, 56 (2013).

\bibitem{hrunski_deposition_2013}
D.~Hrunski et~al.,
\newblock Vacuum {\bf 87}, 114 (2013).

\bibitem{wang_electrode-selective_2016}
J.~K. Wang and E.~V. Johnson,
\newblock Plasma Sources Science and Technology {\bf 26}, 01LT01 (2016).

\bibitem{zhang_control_2015}
Y.~Zhang et~al.,
\newblock Journal of Vacuum Science \& Technology A: Vacuum, Surfaces, and
  Films {\bf 33}, 031302 (2015).

\bibitem{franek_power_2015}
J.~Franek et~al.,
\newblock Review of Scientific Instruments {\bf 86}, 053504 (2015).

\bibitem{dutta_roy_triple_2014}
S.~Dutta~Roy,
\newblock IETE Journal of Education {\bf 55}, 47 (2014).

\bibitem{cauer_verwirklichung_1926}
W.~Cauer,
\newblock Archiv fuer Elektrotechnik {\bf 17}, 355 (1926).

\bibitem{foster_reactance_1924}
R.~M. Foster,
\newblock Bell System Technical Journal {\bf 3}, 259 (1924).

\bibitem{kuo_network_1966}
F.~F.-K. Kuo,
\newblock {\em Network {Analysis} and {Synthesis}},
\newblock John Wiley \& Sons Inc, New York, 2nd revised edition edition, 1966.

\bibitem{mussenbrock_enhancement_2008}
T.~Mussenbrock, R.~P. Brinkmann, M.~A. Lieberman, A.~J. Lichtenberg, and
  E.~Kawamura,
\newblock Physical Review Letters {\bf 101}, 085004 (2008).

\bibitem{mussenbrock_nonlinear_2006}
T.~Mussenbrock, D.~Ziegler, and R.~Peter~Brinkmann,
\newblock Physics of Plasmas {\bf 13}, 083501 (2006).

\bibitem{lieberman_effects_2008}
M.~A. Lieberman, A.~J. Lichtenberg, E.~Kawamura, T.~Mussenbrock, and R.~P.
  Brinkmann,
\newblock Physics of Plasmas {\bf 15}, 063505 (2008).

\bibitem{ziegler_temporal_2009}
D.~Ziegler, T.~Mussenbrock, and R.~P. Brinkmann,
\newblock Physics of Plasmas {\bf 16}, 023503 (2009).

\bibitem{mussenbrock_nonlinear_2007}
T.~Mussenbrock and R.~P. Brinkmann,
\newblock Plasma Sources Science and Technology {\bf 16}, 377 (2007).

\bibitem{schmidt_consistent_2018}
F.~Schmidt, T.~Mussenbrock, and J.~Trieschmann,
\newblock arXiv:1804.05638 [physics]  (2018),
\newblock arXiv: 1804.05638.

\bibitem{vogt_ngspice_2017}
H.~Vogt, M.~Hendrix, and P.~Nenzi,
\newblock Ngspice {Users} {Manual} {Version} 27, 2017,
\newblock Available at: http://ngspice.sourceforge.net/docs/ngspice-manual.
  pdf.

\end{thebibliography}

\end{document}